\documentclass[12pt,onecolumn]{IEEEtran}

\usepackage{amsmath,url}
\usepackage{epsfig,amssymb,amsbsy,verbatim,array,enumerate}
\usepackage{pstricks,psfrag,theorem,cite,paralist,bm}

\let\intern=\iftrue

\def\figref#1{Fig.\,\ref{#1}}%
\def\E{\mathbb{E}}
\def\P{\mathbb{P}}
\def\R{\mathbb{R}}
\def\Z{\mathbb{Z}}
\def\N{\mathbb{N}}

\def\calM{\mathcal{M}}
\def\ie{{\em i.e.}}

\def\sinr{\mathrm{SINR}}

\def\dd{\mathrm{d}}

\def\one{\mathbf{1}}

\def\eps{\varepsilon}
\def\mA{\bm{A}}
\def\mh{\bm{h}}
\def\mm{\bm{m}}

\newtheorem{definition}{Definition}

\newlength{\figwidth}


\begin{document}
\title{Efficient Calculation of Meta Distributions\\and the Performance of User Percentiles} 
\author{Martin Haenggi\\
Dept.~of Electrical Engineering\\
University of Notre Dame
\thanks{%
The support of the U.S.~NSF (grant CCF 1525904) is gratefully acknowledged.}
\vspace*{-4mm}}
\maketitle
\begin{abstract}
Meta distributions (MDs) are refined performance metrics in wireless networks modeled using
point processes.
While there is no known method to directly calculate MDs,
the moments of the underlying conditional distributions (given the point process)
can often be expressed in exact analytical form.
The problem of finding the MD given the moments has several solutions,
but the standard approaches are inefficient and sensitive to the choices of
a number of parameters. Here we propose and explore the use of a method
based on binomial mixtures, which has several key advantages over other methods,
since it is based on a simple linear transform of the moments.
\end{abstract}
\section{Introduction}
{\em Meta distributions.}
A meta distribution (MD) is the distribution of a conditional probability conditioned on the
underlying point process(es) $\Phi$, \ie, the distribution of 
\begin{equation}
   P_t=\P(X>t\mid\Phi) ,
   \label{MD}
\end{equation}
where $X$ is a metric of interest, such as the SIR, SINR, rate, or energy.
Hence $P_t$ is averaged over all randomness in the network except for the spatial locations
of the transceivers (fading, channel access, etc.).
Upon averaging over the point process, the MD
of $X$ at the typical link or user is the cumulative distribution function (cdf)
\[ F(x)=\P(P_t\leq x), \]
with the understanding that $F$ is a function of $t$ also. Depending on the network model,
the probability measure needs to be replaced by the Palm measure given that a transmitter
or receiver resides at a given location. For ergodic point processes $\Phi$, $1-F(x)$ is the fraction
of links or users that achieve $X>t$ with probability at least $x$ in each realization of $\Phi$.
Hence a key advantage of 
the MD over the commonly used mean metric $\P(X>t)=\int_0^1 (1-F(x))\dd x$ is that it 
reveals the performance of user percentiles. For example, $F^{-1}(0.05)$ gives the reliability $x$
that $95\%$ of the users achieve but $5\%$ do not, \ie, the reliability of the ``5\% user".

The MD for the SIR was introduced in \cite{net:Haenggi16twc} and evaluated for two basic
Poisson network models. This refined SIR analysis was extended to cellular networks with D2D underlay in
\cite{net:Salehi17tcom}, to base station cooperation in
\cite{net:Cui18tcom}, to power control for up- and downlink in \cite{net:Wang18tcom},
and to non-orthogonal multiple acesss (NOMA) in \cite{net:Salehi18arxiv}, and to networks with
interference cancellation capability in \cite{net:Wang18wcl}.
Further,
\cite{net:Deng17tcom} extended the SIR to the SINR for mm-wave D2D networks and introduced
the MD of the transmission rate, where $X=W\log_2(1+\sinr)$ for bandwidth $W$, and
\cite{net:Coon18wcl} explores the MD of the secrecy rate. Lastly,
\cite{net:Deng18jsac2} defined the energy MD for a wireless network with energy harvesting capabilities.

{\em Calculation of the meta distributions.}
Since there is no known way to calculate the MD directly, a ``detour" is needed via the moments
of $P_t$. With the moments in hand, the calculation of the MD is an instance of the 
{\em Hausdorff moment problem} \cite{net:Hausdorff23} since the support of $P_t$ is bounded to $[0,1]$.
The standard method to calculate the MD is to use the Gil-Pelaez theorem \cite{net:Gil-Pelaez51},
which requires the integration of the imaginary part of $e^{-\mathrm{i}t\log x}M_{\mathrm{i}t}/t$ over $t\in \R^+$ for
each value of $x$ and $t$,
where $\mathrm{i}^2=-1$ and $M_b=\E(P_t^b)$.
By its nature, the Gil-Pelaez approach requires a careful selection of the range of the numerical
integration (depending on the rate of convergence of the integrand)
and its step size.
Moreover, there is no simple way to bound the error of the resulting approximation.

Another approach is to use only the first and second moments and use the corresponding
beta distribution as an approximation. This method has proven surprisingly accurate but
has its limitations if the actual distribution falls outside the class of beta distributions.

Very recently, the work \cite{net:Guru18arxiv} proposed to use Fourier-Jacobi expansion
to express the MD as an infinite sum of shifted Jacobi polynomials. Truncations
to a finite sum yield approximations, such as the beta approximation above.
The method is promising but its convergence properties are unclear.

Here we propose a method that is appealing due to its simplicity and uniform convergence
properties. It requires the choice of only a single parameter $n$, which denotes the number of
points in $[0,1]$ where the MD is approximated. The approximation is then obtained
by a simple $(n+1)\times(n+1)$ linear transform of the (positive) integer moments,
where the transform matrix is triangular, integer-valued, and only depends on $n$.

\smallskip
{\em Notation.}
$[n]\triangleq \{1,2,\ldots,n\}$, $[n]_0\triangleq \{0\}\cup[n]$.
$\N$ is the set of (positive) natural numbers, and $\N_0=\{0\}\cup \N$.

\section{The Binomial Mixture Method}
For a cdf $F$ with bounded support $[0,T]$, let $\calM$ be the operator that yields the moments
\[ M_n=(\calM F)_n \triangleq \int_0^T x^n \dd F(x), \quad n\in\N , \]
with $M_0=1$.
The Hausdorff moment problem \cite{net:Hausdorff23} is to retrieve $F$ from $M=(M_n)_{n\in\N}$, \ie, to find $\calM^{-1}$. Here we
focus on $T=1$, since the random variables of interest are conditional probabilities. The map $\calM^{-1}$
is unique since $T$ is bounded.

Necessarily, the sequence of moments $M$ of any distribution on $[0,1]$ is completely monotonic, \ie,
\begin{equation}
  (-1)^k(\Delta^k M)_n \geq 0 ,
  \label{diff0}
\end{equation}
where $\Delta^k$ is the iterated difference operator, with $(\Delta M)_n=(\Delta^1 M)_n=M_{n+1}-M_n$.
For example,
$(\Delta^3 M)_3=M_6-3M_5+3M_4-M_3$.
\eqref{diff0} follows from
\[ (-1)^k(\Delta^k M)_n=\int_0^1 x^n (1-x)^k \dd F(x) ,\]
where the right hand side is non-negative.

Our approach is based on the piecewise approximation of $F$ proposed in 
\cite{net:Mnatsakanov03mms}. It is defined as follows.
\begin{definition}[Piecewise approximation]
For any $n\in\N_0$, define the approximate cdf $F_n$ 
\begin{align}
   F_n(x)&
     \triangleq \sum_{k=0}^{\lfloor nx\rfloor}\sum_{j=k}^n \binom nj \binom jk (-1)^{j-k} M_j, \quad x\in (0,1],
   \label{fn}
\end{align}
and $F_n(0)=0$,
where $\lfloor u\rfloor$ is the largest integer smaller than or equal to $u$. 
\end{definition}
The key property of this approximation is that $F_n(x)\to F(x)$ as $n\to\infty$ for each $x$ at which $F$ is continuous
\cite{net:Mnatsakanov03mms,net:Mnatsakanov08stapro}, \ie, \eqref{fn} converges to the map $\calM^{-1}$.

$F_n$ has $n+1$ discontinuities, uniformly spread at $x=k/n$, $k\in[n]_0$.

\section{Numerical Recipe}
\subsection{Sampling $F_n$}
\begin{figure}
\centerline{\epsfig{file=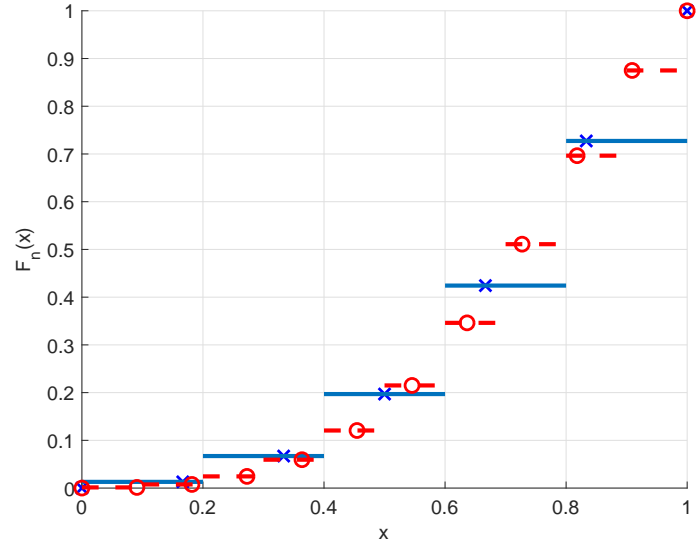,width=\figwidth}}
\caption{$F_5(x)$ (solid) and $F_{10}(x)$ (dashed) for a beta distribution with parameters $\alpha=5$ and $\beta=2$. The $\times$ denote
the samples $F_5(k/6)$, $k\in [6]_0$, and the $\circ$ denote the samples $F_{10}(k/11)$, $k\in [11]_0$.}
\label{fig:sampling}
\end{figure}

For the numerical calculation of \eqref{fn}, $x\in[0,1]$ needs to be sampled at discrete values.
We choose $x_k=k/(n+1)$, for $k\in[n+1]_0$, which is the densest 
uniform sampling such that $F_n(x_k)\neq F_n(x_j)$, $i\neq j$. Moreover, if $F$ is the cdf of a
uniform random variable, the linear interpolation of $F_n(x_k)$ gives the exact cdf for any $n\in\N_0$.
\figref{fig:sampling} gives an example of $F_5$ and $F_{10}$ for a beta distribution and shows
the sampling values.

\subsection{Approximate Probability Density Function}
Letting
\begin{equation}
   h_k\triangleq \sum_{j=k}^n \binom nj \binom jk (-1)^{j-k} M_j, \quad k\in[n]_0,
   \label{hk}
\end{equation}
we have 
\[ F_n(x_k)=\sum _{m=0}^{k-1} h_m , \quad k\in[n+1]_0,\]
and $f_n(x_k)\triangleq (n+1) h_k$ is an approximation of the probability density function (pdf) since
$F_n(x_k)$ can be interpreted as a Riemann sum
\[ \frac1{n+1}\sum_{m=0}^{k-1} f_n(x_m) ,\]
which converges to $F(x_k)$ as $n\to\infty$.

\begin{figure}
\centerline{\epsfig{file=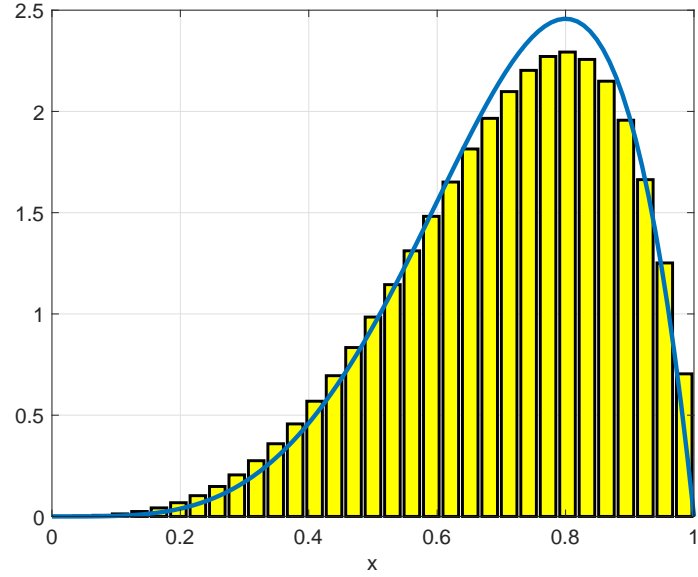,width=\figwidth}}
\caption{$f_{32}(x)$ for a beta distribution with parameters $\alpha=5$ and $\beta=2$.
The solid curve is the exact beta pdf.}
\label{fig:histo}
\end{figure}

\figref{fig:histo} shows an exact beta pdf and the approximation $f_{32}$.

\subsection{The Method as a Linear Transform}
Eqn.~\eqref{hk} shows that $h_k$, $k\in [n]_0$, is just a linear combination
of the moments $M_j$, $j\in[n]_0$. Accordingly, we can write $\mh=(h_k)_{k=0}^n$
as a {\em linear transformation} of $\mm=(M_j)_{j=0}^n$, \ie,
\begin{equation}
   \mh=\mA \mm,
   \label{linear}
\end{equation}
where $\mm$ and $\mh$ are understood as column vectors and
the transform matrix $\mA\in\Z^{(n+1)\times(n+1)}$
is, from \eqref{hk}, given by
\[ A_{ij}\triangleq\binom nj \binom ji (-1)^{j-i}\one(j\geq i),\quad i,j\in [n]_0,\]
where $\one$ is the indicator function.
For example, for $n=4$,
\[ \mA=\begin{bmatrix}
       1  &  -4  &   6  &  -4  &   1\\
     0  &   4 &  -12   & 12 &   -4\\
     0 &    0  &   6  & -12  &   6\\
     0  &   0  &   0 &    4  &  -4\\
     0  &   0  &   0  &   0   &  1
     \end{bmatrix}
     \]
In addition to being (right) upper triangular, $\mA$ is also symmetric with respect to the 
antidiagonal. Hence only about $n^2/4$ entries need to be calculated\footnote{The
exact number if $[(n+2)^2-\one(n\text{ odd})]/4$.}.

The key benefit of the linear mapping \eqref{linear} is that the transform matrix
needs to be calculated only once for the desired level of accuracy $n$. This can
be done ``offline", which reduces the calculation of the approximate meta distribution
to a simple matrix-vector multiplication requiring $(n^2+3n)/2$ multiplications
(once the moments are known).

\subsection{Choosing the Required Accuracy}
The default accuracy of standard mathematical software such as Matlab$^\text{\textregistered}$
is insufficient to calculate $F_n$ for $n$ larger than about 30. Here we
give an estimate of the number of decimal digits $b$ that enable the
calculation $F_n$ with sufficient precision, for arbitrary $n$.

The maximum of $|\mA|$ occurs on the antidiagonal
$\bm{d}=(A_{0n},A_{1,n-1},\ldots,A_{n0})$. Indexing $\bm{d}$ from $0$ to $n$,
 $d_{\lceil n/3 \rfloor}$ is the largest entry in absolute value, where $\lceil u \rfloor$
is rounding $u$ to the nearest integer.
Stirling's approximation yields
\[ \max|\mA| \sim \frac{\sqrt{27}}{2\pi} \frac{3^n}{n} ,\quad n\to\infty.\]
With $\sqrt{27}/(2\pi)<1$ and $\log_{10}(3)<1/2$, about $n/2-\log_{10}n$
decimal digits are needed to calculate $\mA$.

For the moment vector $\mm$, it depends how quickly the moments  go to zero.
The two extreme cases are the uniform distribution, where $M_n=1/(n+1)$, and the
(degenerate) step function, where $M_n=\nu^n$, where $\nu<1$ is the constant value of
the random variable. This exponential decay of the moments is not of
practical interest, as a comparison of $M_1^2$ and $M_2$ would immediately reveal
that the distribution is degenerate. 

This leaves two qualitatively different cases of interest.
\subsubsection{Case 1: Superpolynomial decay}
In this case, $-\log M_n/\log n\to\infty$ but there exists $c>0$ and
$0<\delta<1$ such that 
\[ M_n \leq 10^{-c n^\delta} ,\quad\forall n\geq 0.\]
As a result,
\[ b=n/2+cn^\delta-\log_{10} n \]
decimal digits are sufficient.

\subsubsection{Case 2: Polynomial decay}
In this case, there exists $s\geq 1$ such that
\[ M_n \leq (n+1)^{-s},\quad\forall n\geq 0 ,\]
and 
\[ b=n/2+(s-1)\log_{10} n \]
decimal digits suffice.

The parameters $c$, $\delta$, or $s$ can be estimated by inspection
of the moment sequence. As a simple rule of thumb, 
$b=n/2+16$ is a sensible choice. Assuming $c=1$ and $\delta=1/2$
in the superpolynomial case, this is sufficient for $n$ up to $350$.
In the polynomial case, $s=6$ can be accommodated up to $n=1000$, which
is likely to be sufficient for all practical purposes.

\section{Accuracy of the Method}
A detailed study of the convergence properties
of $F_n$ to $F$ is provided in \cite[Theorem 2]{net:Mnatsakanov08stapro}.
Letting $f(x)=F'(x)=\dd F(x)/\dd x$ be the exact pdf and $f'(x)$ its derivative, the theorem
asserts that as $n\to\infty$,
\begin{equation}
   \|F_n-F\| \leq \frac{1}{n+1}\left(\|f\|+\frac{\|f'\|}{2}\right) + o(n^{-1}) ,
   \label{acc}
\end{equation}
where
$\|f\|=\sup_{x\in[0,1]} |f(x)|$. So if $f'$ is bounded, convergence is uniform, and 
the maximum error decreases with $1/n$. 

\begin{figure}
\centerline{\epsfig{file=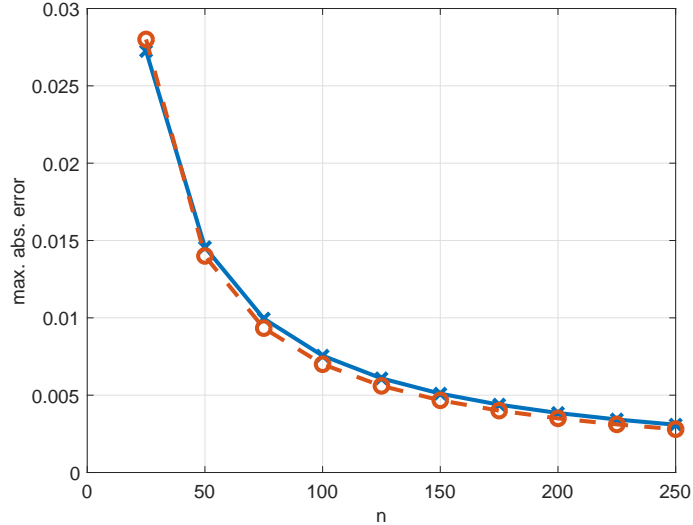,width=\figwidth}}
\caption{The solid line shows the maximum error $\max_{k\in [n]}|F_n(x_k)-F(x_k)|$ as a function of $n$ for a beta distribution with parameters $\alpha=5$ and $\beta=2$.
The dashed line shows $0.7 n^{-1}$.}
\label{fig:max_error}
\end{figure}

\figref{fig:max_error} shows how the maximum error decreases with $n$ in the case of 
a beta distribution for $\alpha=5$, $\beta=2$, for which $\|f\|+\|f'\|/2=10911/625\approx 17.5$.
The pre-constant $0.7$ in the reference curve in the figure is about 25 times smaller than that,
indicating that the bound is conservative in some cases.
The figure also shows that the $1/n$ scaling of the maximum error starts at modest $n$ already.

For the beta distribution with $\alpha<1$ or $\beta<1$, the $1/n$ scaling does not
hold since $\|f'\|$ is not bounded.

\begin{figure}
\centerline{\epsfig{file=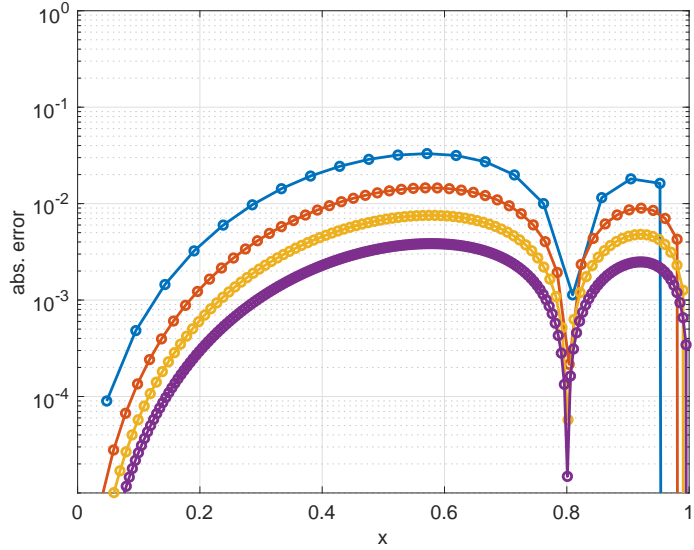,width=\figwidth}}
\caption{Absolute error $|F_n(x_k)-F(x_k)|$ for $n=20,50,100,200$ for a beta distribution with parameters $\alpha=5$ and $\beta=2$.}
\label{fig:abs_error}
\end{figure}

\section{Rate-reliability Trade-off for User Percentiles}
As an application, we explore the rate-reliability trade-offs for users that are in a
certain percentile of all users. 
In this case, the underlying random variable $X$ in \eqref{MD} is the 
SIR, and $t$ is the SIR threshold required for successful reception, henceforth
denoted by $\theta$. We are interested in the pairs $(\theta,x)$ for which
$F_n(x)=p$, where $p$ is the user percentile. For example, setting $p=0.05$,
choosing $\theta>0$ and solving for $x$ yields the $(\theta,x)$ pairs the 5\% user
achieves.

\begin{figure}
\hfill{\epsfig{file=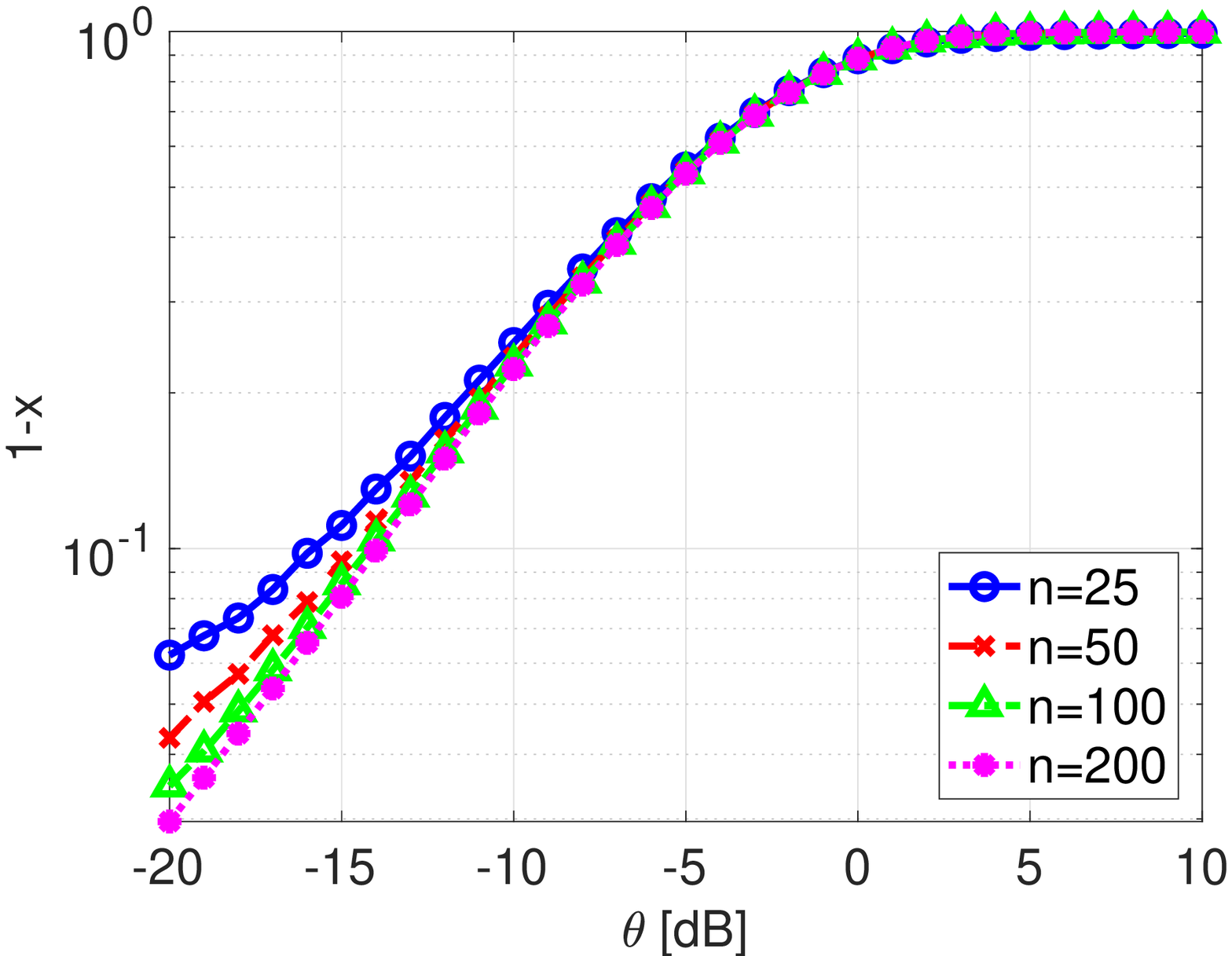,width=.9\figwidth}}\hfill
{\epsfig{file=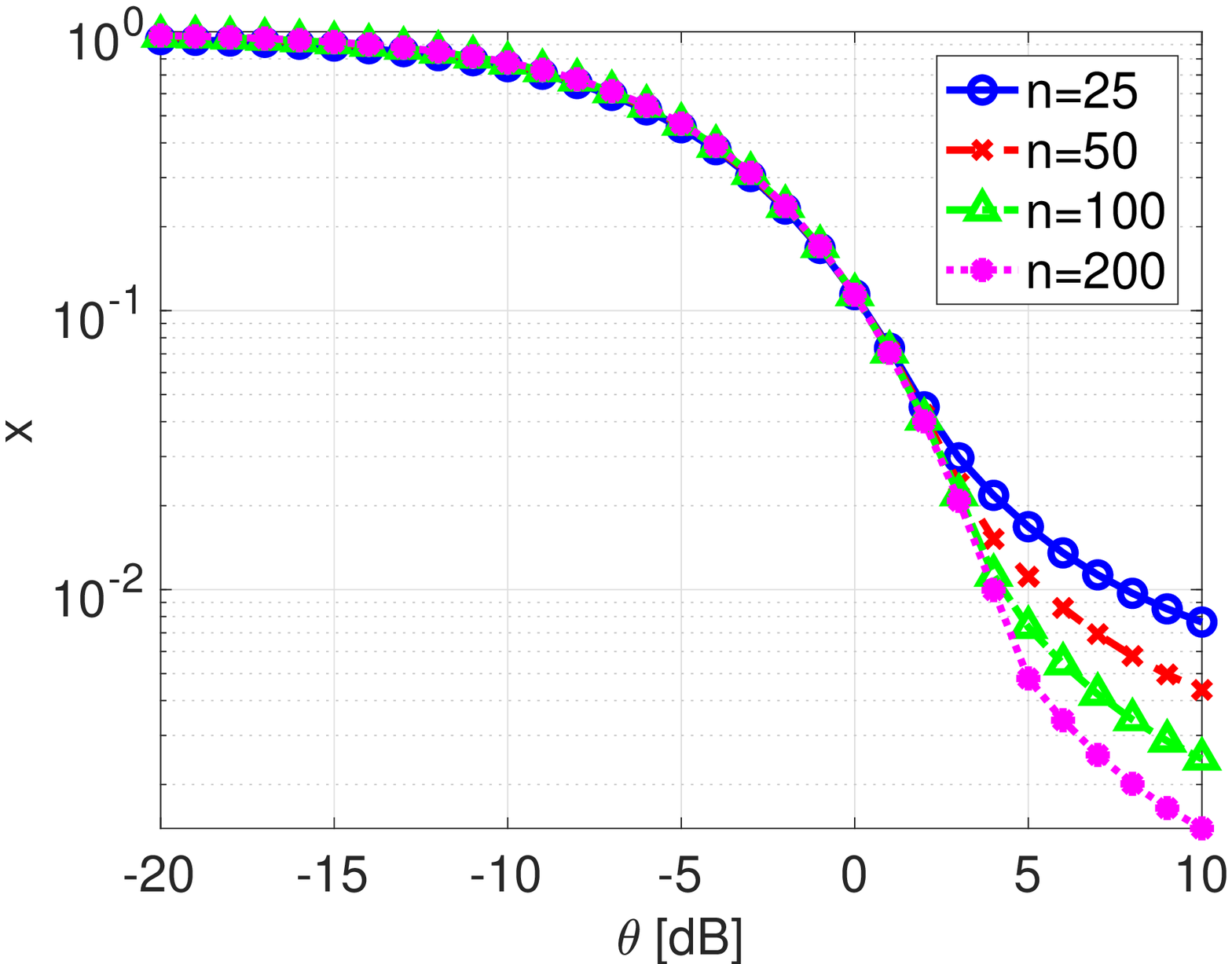,width=.9\figwidth}}\hfill
\caption{Pairs $(\theta,1-x)$ (left plot) and $(\theta,x)$ (right plot) for which $F_n(x)=0.1$ for $n\in\{25,50,100,200\}$.}
\label{fig:user_conv}
\end{figure}

We use the standard downlink Poisson cellular model
with nearest-base station association, power-law path loss and Rayleigh fading,
for which the moments of the conditional success probability are given in 
\cite{net:Haenggi16twc} as
\[ M_n=(\,_2F_1(n,-\delta;1-\delta,-\theta)^{-1} ,\]
where $_2F_1$ is the Gauss hypergeometric function and $2/\delta$ is the path loss
exponent.
\figref{fig:user_conv} explores the effect of $n$ on the trade-off between $\theta$ and $x$
for the 10\% user.
As can be seen, for $\theta\in [-10,0]$ dB, $n=25$ is
sufficiently accurate. For $\theta<-10$ dB, the reliability $x$ increases with $n$,
indicating that all curves are (increasingly tight) lower bounds, while for $\theta>10$ dB, the
reliability decreases with $n$, \ie, the curves are upper bounds.

\begin{figure}
\hfill{\epsfig{file=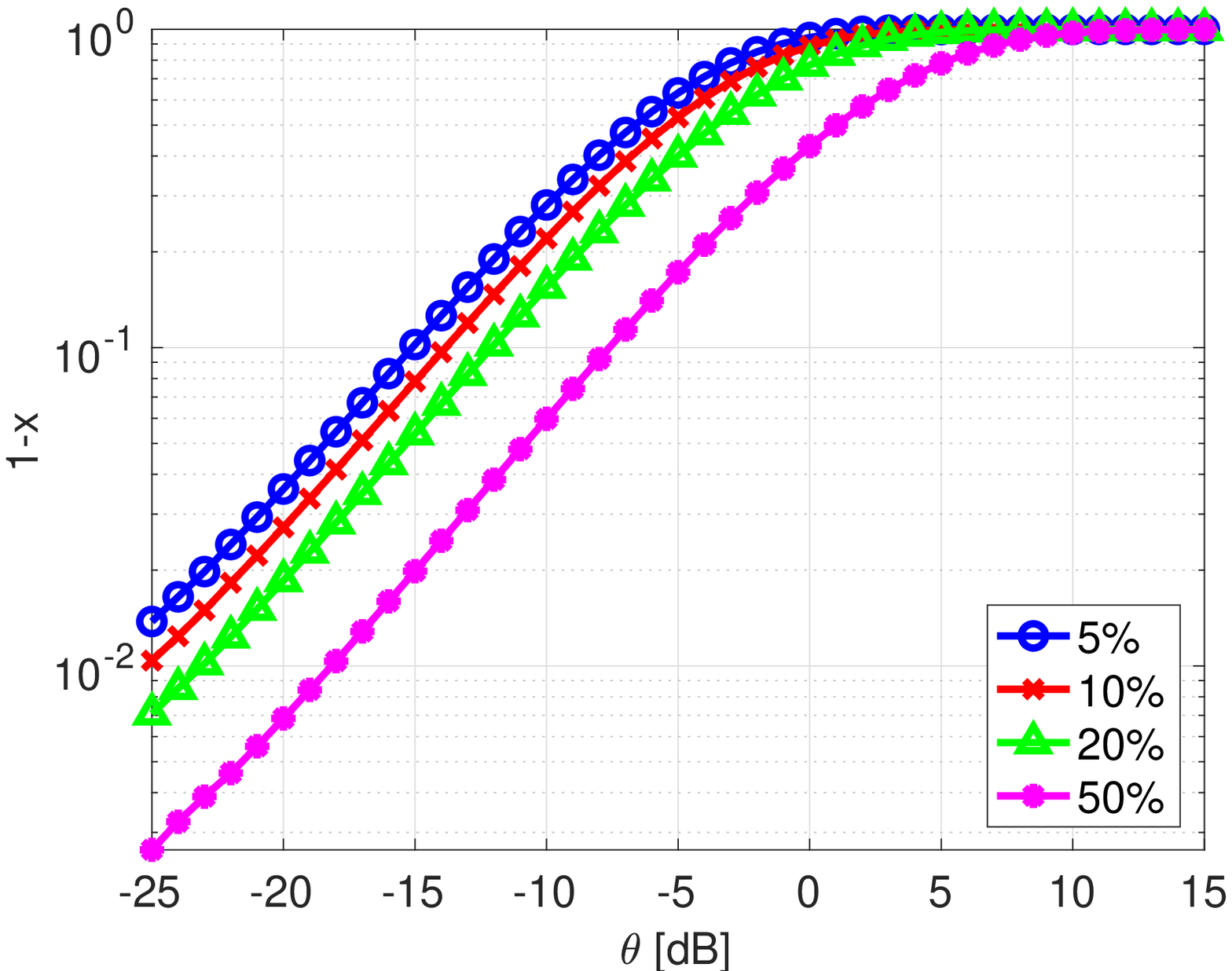,width=.9\figwidth}}\hfill
{\epsfig{file=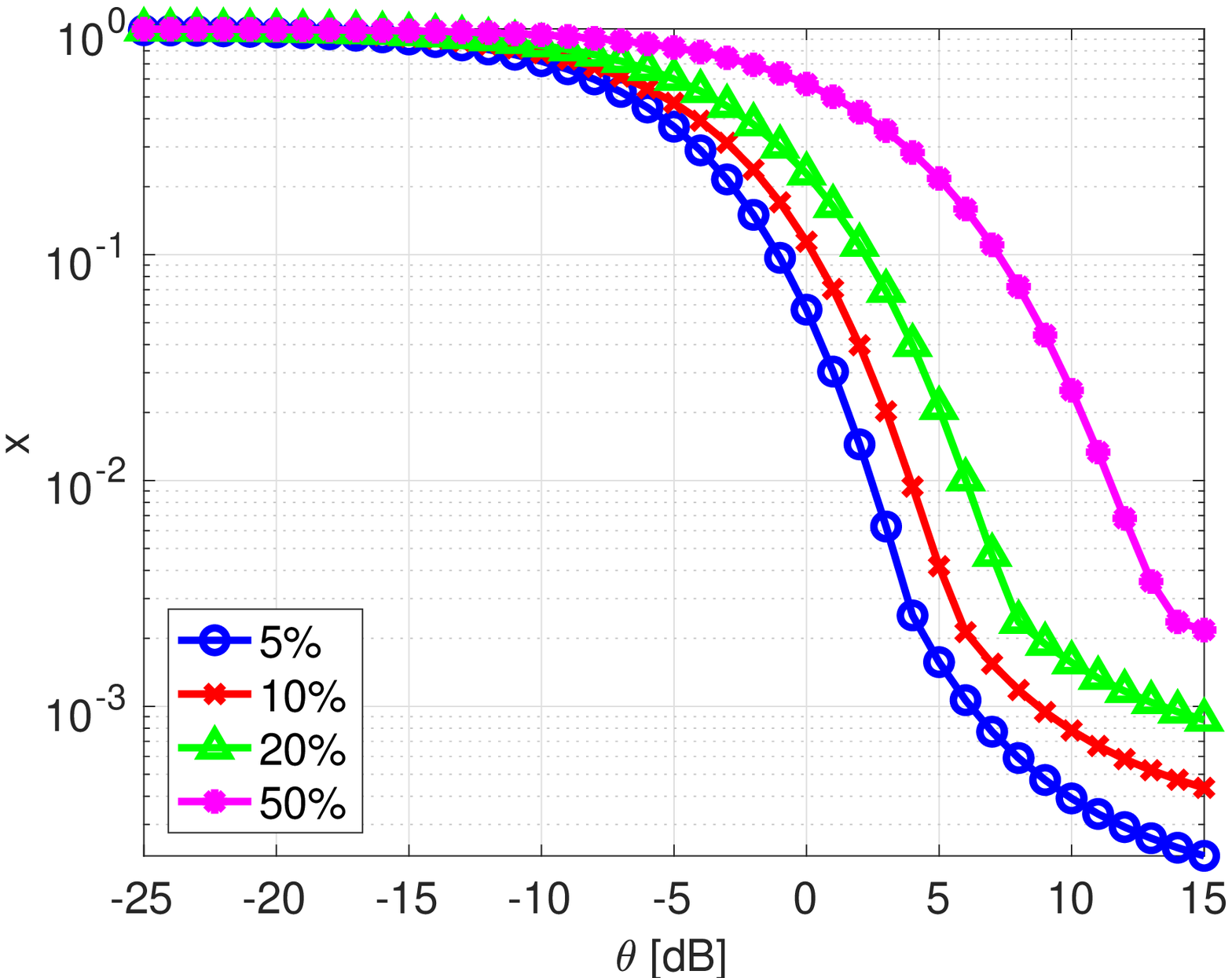,width=.9\figwidth}}\hfill
\caption{Pairs $(\theta,1-x)$ (left plot) and $(\theta,x)$ (right plot) for which $F_n(x)=0.05, 0.1, 0.2, 0.5$ for $n=400$.}
\label{fig:perc}
\end{figure}

\begin{figure}
\centerline{\epsfig{file=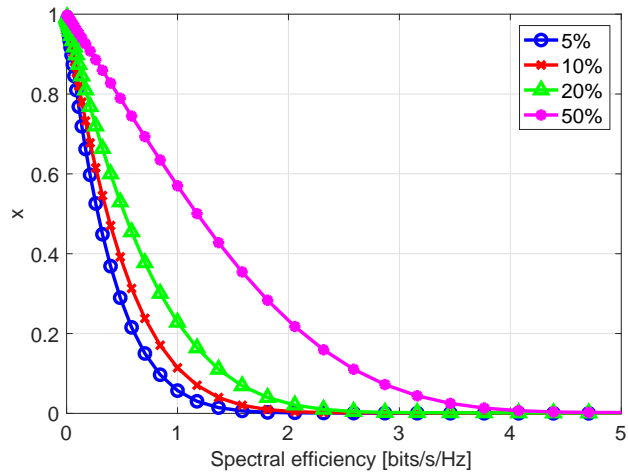,width=.9\figwidth}}
\caption{Rate-reliability trade-off (linear scales) for user percentiles.}
\label{fig:rrt}
\end{figure}

In \figref{fig:perc}, the $(\theta,x)$ pairs for the 5\%, 10\%, 20\%, and 50\% user
percentiles are shown,
and \figref{fig:rrt} presents the spectral efficiency-reliability trade-off $(\log_2(1+\theta),x)$ for the same percentiles, both for $n=400$.
These figures involve the evaluation of more than 16,000 values of the MD ($n=400$ times
$41$ values of $\theta$), which takes less than 20 seconds with Matlab$^\text{\textregistered}$
on a standard desktop computer
(not including the time to calculate the transform matrix $\mA$, which is done only once, and the
moments for each value of $\theta$).

Interestingly, the curves in \figref{fig:perc} for the different percentiles run parallel asymptotically for
both $\theta\to 0$ and $\theta\to\infty$. Moreover, the gap is approximately the same
on both sides. Between the 5\% and the 50\% users, it is about 10 dB. This asymptotic gap
can be used to quantify the fairness between users.

\section{Conclusions}
The binary mixture method is extremely simple and very efficient for the numerical computation
of meta distributions, because it is a linear transform of the moment sequence.
Only a single parameter needs to be chosen, namely
the number of points $n$ to be calculated. The approximated cdf $F_n$ converges
uniformly to the exact $F$ at a rate of $1/n$, and the required number of decimal
digits is about $n/2$.

\bibliographystyle{IEEEtr}

 
 \end{document}